\documentclass[prd,aps,twocolumn,preprintnumbers, showpacs, nofootinbib,superscriptaddress,notitlepage]{revtex4-1}
\usepackage{mathrsfs}
\usepackage{amsfonts}
\usepackage{amsmath}
\usepackage{slashed}
\usepackage{array}
\usepackage{verbatim}
\usepackage{epsfig}
\usepackage{graphicx}
\usepackage{color}
\usepackage[dvipsnames]{xcolor}

\definecolor{red}{rgb}{1,0,0}

\newcommand{\beq}{\begin{eqnarray}}
\newcommand{\eeq}{\end{eqnarray}}

\newcommand{\tr}{{\rm tr}  }

\def\be{\begin{equation}}
\def\ee{\end{equation}}
\def\bea{\begin{eqnarray}}
\def\eea{\end{eqnarray}}

\begin{document}

\title{Spontaneous chiral symmetry breaking and mass gap of QCD in finite volume}


\author{
Xiaolan Meng}
\affiliation{University of Chinese Academy of Sciences, School of Physical Sciences, Beijing 100049, China}
\affiliation{CAS Key Laboratory of Theoretical Physics, Institute of Theoretical Physics, Chinese Academy of Sciences, Beijing 100190, China}

\author{
Bolun Hu}
\affiliation{CAS Key Laboratory of Theoretical Physics, Institute of Theoretical Physics, Chinese Academy of Sciences, Beijing 100190, China}

\author{Yi-Bo Yang}
\affiliation{University of Chinese Academy of Sciences, School of Physical Sciences, Beijing 100049, China}
\affiliation{CAS Key Laboratory of Theoretical Physics, Institute of Theoretical Physics, Chinese Academy of Sciences, Beijing 100190, China}
\affiliation{School of Fundamental Physics and Mathematical Sciences, Hangzhou Institute for Advanced Study, UCAS, Hangzhou 310024, China}
\affiliation{International Centre for Theoretical Physics Asia-Pacific, Beijing/Hangzhou, China}

\date{\today}

\begin{abstract}
We present the lattice QCD simulation with the 2+1+1 flavor full QCD ensembles using near-physical quark masses and different spatial sizes $L$, at $a\sim$ 0.055 \;fm. The results suggest that chiral symmetry is effectively restored at $L\le 1.0$ \;fm, while the gap between the nucleon and pion masses remains.
\end{abstract}

\maketitle

\section{Introduction}

Both spontaneous chiral symmetry breaking and confinement are intrinsic features of the strong interaction between quarks and gluons, although their relationship is the subject of debate. Lattice QCD has confirmed the restoration of chiral symmetry and certain aspects of de-confinement at high temperatures~\cite{Bazavov:2011nk}. However, there are also studies arguing that the strong interaction between quarks still exists, and therefore, deconfinement is incomplete (e.g., the recent $N_f$=2 study~\cite{Rohrhofer:2019qwq}). These discussions are based on the spatial correlation function of meson interpolators, which differs from the standard hadron mass and would be sensitive to flavors (or other systematics)~\cite{Chiu:2023hnm}. 

All the above lattice QCD calculations focus on the infinite volume limit $m_{\pi}L\gg 1$, as the finite volume effect on the physical observable can be significant when $m_{\pi}L$ is not large enough~\cite{Gasser:1987ah}. As evidence, in a Lattice QCD simulation at $a\sim$0.11 \;fm with $m_{\pi}L\sim 1$ using near physical light quark mass and $L\sim$ 1.8 \;fm, an effective restoration of chiral symmetry in the mesonic two-point functions was observed~\cite{Fukaya:2007pn}. A system with $m_{\pi}L\le 1$ is referred to as the $\epsilon$-regime of the chiral perturbative theory ($\chi$PT)~\cite{Gasser:1987ah}, and reasonable leading order low energy constants of $\chi$PT can be extracted using the above lattice calculations~\cite{Fukaya:2007pn}.

The finite temperature Lattice QCD simulation requires a smaller temporal size to reach higher temperatures. Therefore, one cannot extract the standard ground state hadron mass based on the long-distance temperature correlation function, and the use of the spatial correlation function would be unavoidable. On the other hand, chiral symmetry restoration can also be observed with a long enough temporal size in the $\epsilon$-regime, allowing for reliable extraction of the ground state hadron mass. In this work, we present a preliminary study on the masses of iso-vector mesons and nucleon at different $L$, using the clover fermion on the HISQ ensembles with near-physical quark masses. We find that the mass gap between the nucleon and pion masses remains even when $L$ is as small as 0.2 \;fm.

\section{Numerical setup}

\begin{figure}[thb]
    \includegraphics[width=0.45\textwidth]{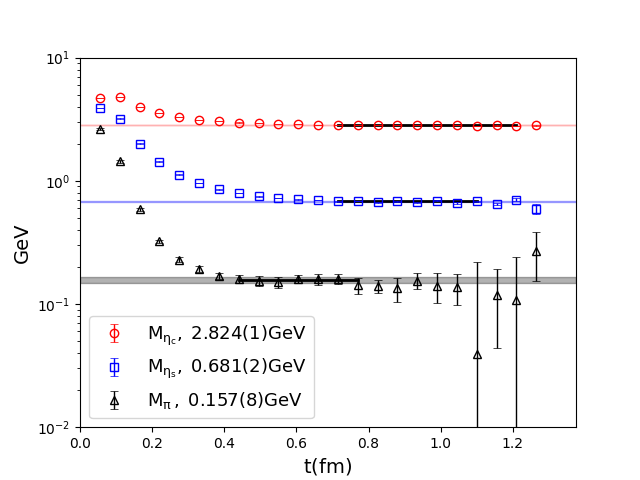}
    \caption{The pion, $\eta_s$ $\eta_c$ effective masses on the $48^4$ lattice at $a\sim0.055$ \;fm, using the HISQ action.}
    \label{fig:ps_masses}
\end{figure}

In order to preserve the chiral symmetry of the light quark in the sea, we chose the HISQ fermion action and one-loop improved Symanzik gauge action used by the MILC collaboration to generate the configurations. The bare coupling and quark masses were interpolated to those at $a=0.052$ \;fm based on the parameters used by the MILC configurations~\cite{MILC:2010pul,MILC:2012znn}, as such a lattice spacing is used by the CLQCD ensemble H48P32~\cite{Hu:2023jet}. The lattice spacing determined through the Wilson flow~\cite{Luscher:2010iy} is approximately 0.055 \;fm and close to the target value.

The corresponding effective mass of the pion, $\eta_s$, and $\eta_c$ using this lattice spacing are shown in Fig.~\ref{fig:ps_masses}. Based on the constant fit at a relatively large $t$, we can extract $m_{\pi}=0.157(8)$ GeV, which is quite close to the experimental value 0.135 GeV. The $\eta_s$ mass $m_{\eta_s}=0.681(2)$ GeV is also close to the most precise lattice determination of 0.68963(18) MeV~\cite{Borsanyi:2020mff}, and the $\eta_c$ mass $m_{\eta_c}=2.824(1)$ GeV is only 5\% lower than the experimental value of 2.98 GeV~\cite{ParticleDataGroup:2020ssz}.

Since the mixed action effects have been shown to be ${\cal O}(a^4)$ with the HISQ fermion sea~\cite{Zhao:2022ooq}, we use the tadpole improved clover fermion action with the HYP smeared gauge field for the valence quark, and tune the corresponding pion mass to be about 230 MeV  on the $48^4$ lattice, which is close to the unitary pion mass. Such a choice is more convenient for investigating the meson and nucleon spectrum without the additional difficulty from the taste-breaking effect of the HISQ action.

\begin{table}[ht!]                   
\caption{}  
\begin{tabular}{c c c c | c c}              
$10/g^2$ & $m_la$ & $m_sa$ & $m_ca$ & $V$ & $n_{\rm cfg}$ \\
\hline 
6.784 & 0.000731 & 0.01975 & 0.2293 & $4^3\times 96$ & 238\\ 
& & & & $8^3\times 96$ & 350\\
& & & & $8^2\times 32\times 96$ & 100\\
& & & & $10^3\times 96$ & 148\\
& & & & $12^3\times 96$ & 150\\
& & & & $16^3\times 96$ & 146\\
& & & & $20^3\times 96$ & 148\\
& & & & $24^3\times 96$ & 151\\
& & & & $32^3\times 96$ & 98\\
& & & & $40^3\times 96$ & 72\\
& & & & $40^3\times 96$ & 80\\
& & & & $48^3\times 48$ & 100\\
\hline
\end{tabular}  
\label{tab:ensem}
\end{table}

The ensembles we generated for the this work are summarized in Table~\ref{tab:ensem}. 

\section{Results}

We calculate the hadron temporal two points correlators including pseudo scalar(P), scalar(S) and proton(N) with clover valence mass  $m_{u}=-0.0414$  at near physical point to extract hadron masses according to the formula 
\begin{equation}
\mathcal{C}_{P,S}(t)= \sum_{\vec{x},\vec{y}} \langle \mathcal{O}(\vec{x},t)\mathcal{O}(\vec{y},0)^{\dagger} \rangle,
\end{equation}
with P,S interpolators defined as
\begin{equation}
 \mathcal{O}_P = \bar{u} \gamma_{5}d, \;
 \mathcal{O}_S = \bar{u} d,
\end{equation}
and nucleon two-point functions defined as
\begin{align}
\mathcal{C}_{N}(t)=&\sum_{\vec{x},\vec{y}} \epsilon^{abc}\epsilon^{a'b'c'} \langle
\bar{d}^{a}(\vec{x},t)(C\gamma_{5})\bar{u}^{b^{T}}(\vec{x},t)\bar{u}^{c}(\vec{x},t)  \notag  \\
& \Gamma_e u^{a'}(\vec{y},0)d^{b'^{T}}(\vec{y},0)(C\gamma_{5})u^{c'}\rangle,
\end{align}
with $\Gamma_e=(1+\gamma_4)/2$.
When computing quark propagators, Coulomb wall sources with random positions in the temporal direction are used to reduce autocorrelation effects between configurations. The values for the pion masses are extracted from fitting correlators by one-state fit with fitting range selected to ensure acceptable $\chi^{2}$/d.o.f.$\sim$1,
\begin{equation}
\mathcal{C}_{P}(t)=A\ {\rm cosh}(m_{\pi}(T-t))
\end{equation}
\begin{figure}[thb]
    \includegraphics[width=0.45\textwidth]{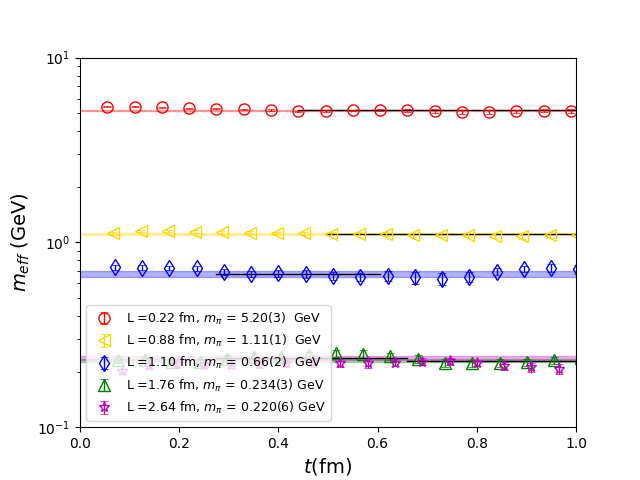}
    \caption{The non-singlet pion effective masses at different spatial size $L$.}
    \label{fig:pion_eff_masses_L}
\end{figure}
with fitting parameters A, $m_{\pi}$. The result is shown in Fig.~\ref{fig:pion_eff_masses_L}. We observe clear plateaus in the effective mass $m_{\text{eff}}(t)$ for all values of $L$, which suggests that $\mathcal{C}_{P}(t)$ exhibits exponential decay regardless of the spatial volume when $T$ is large enough.
When $L \sim 0.23$ \;fm, $m_{\pi}$ becomes massive around 5 GeV. The $m_{\pi}$ rapidly decreases as $L$ increases, and stabilizing at around 230 MeV when $L  > 1.8$ \;fm.
\begin{figure}[thb]
    \includegraphics[width=0.45\textwidth]{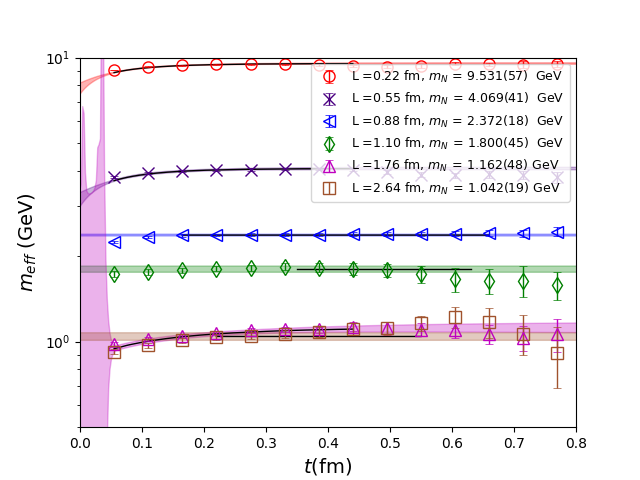}
    \caption{The nucleon effective masses at different spatial size $L$.}
    \label{fig:nuc_eff_masses_L}
\end{figure}

\begin{figure}[thb]
    \includegraphics[width=0.45\textwidth]{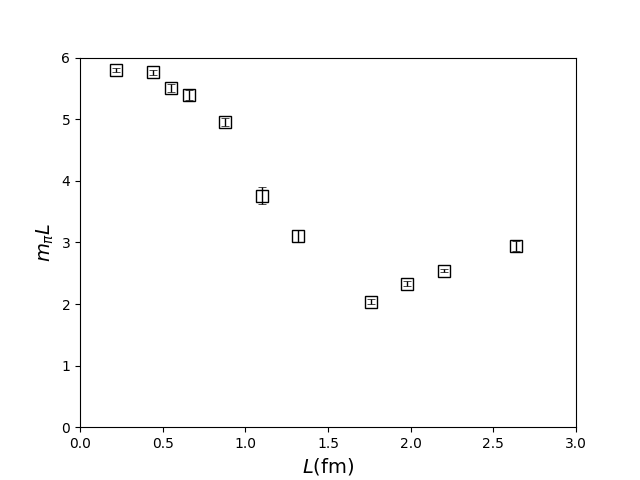}
    \caption{$m_{\pi} L$ as the function of $L$.}
    \label{fig:regime}
\end{figure}

\begin{figure}[thb]
    \includegraphics[width=0.5\textwidth]{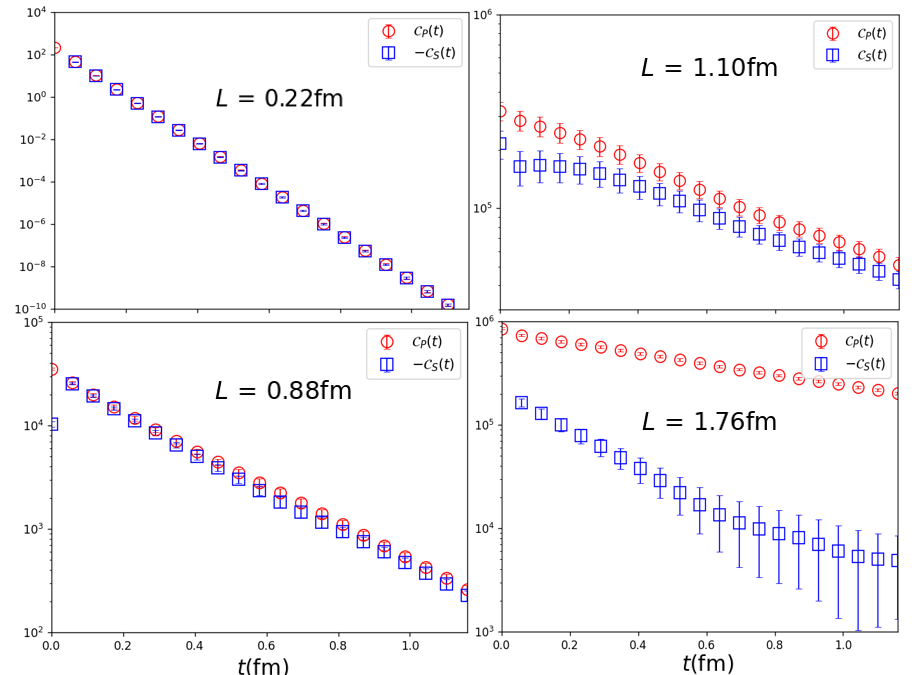}
    \caption{pion and iso-vector scalar meson $a_0$ two-point function as function of $t$, at $L=\{0.2,0.9,1.1,1.8\}$ \;fm. }
    \label{fig:meson_corr}
\end{figure}

\begin{figure}[thb]
    \includegraphics[width=0.45\textwidth]{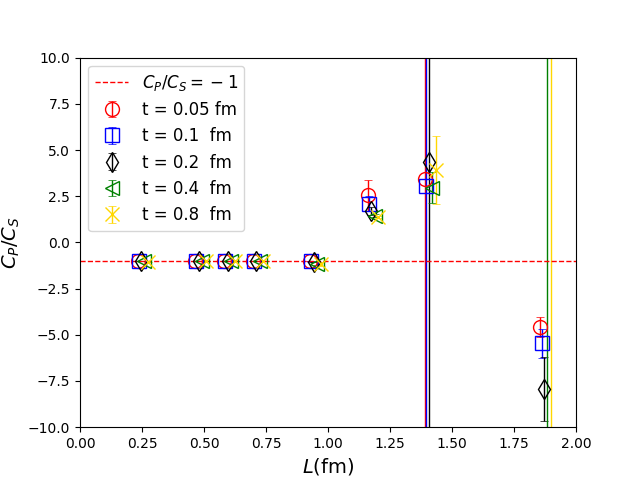}
    \caption{The chiral symmetry breaking effect $C_{P}(t)/C_{S}(t)$ as the function of $L$, at $t=\{0.05, 0.1, 0.2,0.4,0.8\}$ \;fm.}
    \label{fig:chiral_L}
\end{figure}

For proton mass, due to poor signal at large $t$, we employ a two-state fit for $\mathcal{C}_{N}(t)$ at relatively smaller $t$ in the ensembles with certain $L$ to eliminate the excited state contamination,
\begin{equation}
\mathcal{C}_{N}(t) = c_{1}e^{-m_{N}t}(1+c_{2}e^{-\Delta mt}),
\end{equation}
where $m_N$, $\Delta m$ and $c_{1,2}$ are fitting parameters.
The remaining cases are fitted with single state($c_{2} = 0$). It is revealed that proton masses also follow similar patterns with changes in volume as pion: $m_{N}$ decreases when spatial extent increases, and tends to about 1 GeV at infinite volume limit.  
\begin{figure}[thb]
    \includegraphics[width=0.45\textwidth]{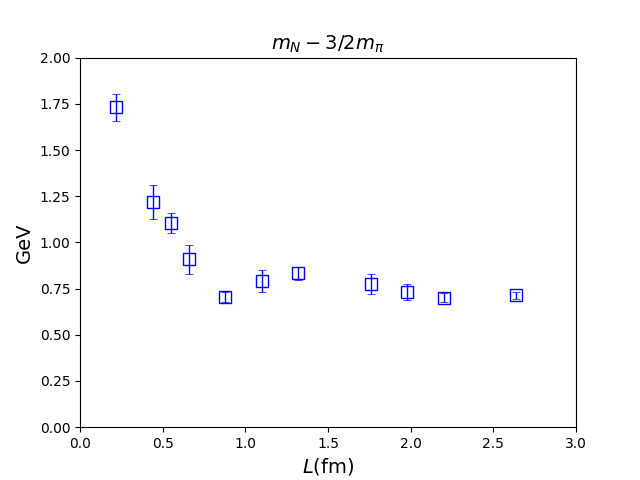}
    \caption{The mass difference $m_N-3/2m_{\pi}$ as function of $L$.}
    \label{fig:hadron_masses_diff_L}
\end{figure}
Since pion and nucleon have different number of quarks, we define the combination $m_{N}-1.5m_{\pi}$ to examine the mass gap between proton and pion based on their quark contents. The result in Fig.~\ref{fig:hadron_masses_diff_L} shows the mass gap between the proton and pion remains above 0.7 GeV and increases even at small volume, which suggests that the QCD confinement remains even at very small $L$. From another perspective shown in Fig.~\ref{fig:regime}, $m_{\pi}L$ has a lower limit(around 2) and tends towards 6$\sim 2\pi$ at small volumes, which imply the epsilon regime cannot be reached simply by reducing the spatial volume and QCD is not completely dominated by the possible massless pion field at small $L$.
In addition, the variation in the pion and iso-vector scalar meson $a_0$ two-point function can reflect the nature of chiral symmetry. If chiral symmetry is effectively restored, the quark propagator $\mathcal{S}(x,y)=\psi(x)\bar{\psi}(y)$ on each individual configurations with negligible quark mass will anti-commutate with $\gamma_5$, $\{ \gamma_{5}, \mathcal{S} \}\sim0$. {Thus }we have
\begin{equation}
\mathcal{C}_{P}(t)= \sum_{\vec{x},\vec{y}} \langle \tr [ \gamma_{5} \mathcal{S} (\vec{0}, 0 ; \vec{y} , t) \gamma_{5} \mathcal{S}(\vec{y} , t ; \vec{0}, 0 )  ]  \rangle = - \, \mathcal{C}_{S}(t),
\end{equation}
and then $\mathcal{C}_{P}(t)$ and $\mathcal{C}_{S}(t)$ will be degenerate. The meson two-point functions are shown in Fig.~\ref{fig:meson_corr} and Fig.~\ref{fig:chiral_L}. When $L < 0.7 {\rm \;fm}$, $\mathcal{C}_{P}(t)=-\mathcal{C}_{S}(t)$ holds effectively for all time slices except $t=0$, which may indicate the effectively restoration of chiral symmetry. When $L$ reaches 0.9 \;fm, $\mathcal{C}_{P}(t)$ begins to slightly deviate from $-\mathcal{C}_{S}(t)$ and significantly deviates as L increases to 1.1 \;fm, including an unexpected sign change in the latter. $\mathcal{C}_{S}(t)$ returns to negative values at $L \sim 1.8$ \;fm and show a faster exponential decay at relatively small $t$ which would corresponds to the physical $a_0$ state in the realistic QCD. $\mathcal{C}_{S}(t)$ exhibit irregular behaviours as volume increase continually, possibly due to the contamination from the other $0^{++}$ states.

\begin{figure}[thb]
    \includegraphics[width=0.45\textwidth]{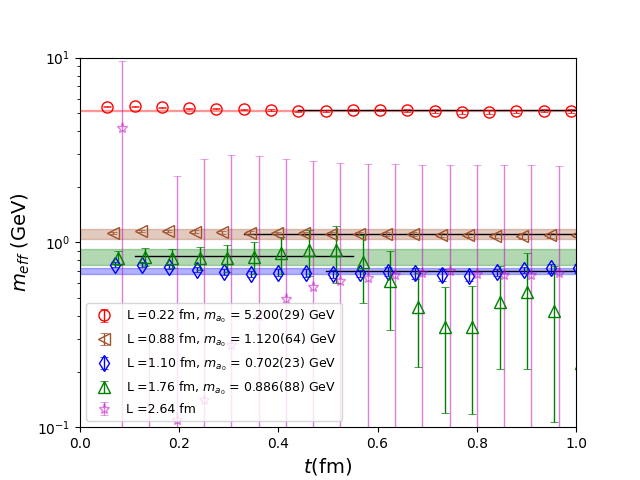}
    \caption{The $a_0$ effective masses at different spatial size $L$. }
    \label{fig:s_eff_masses_L}
\end{figure}
\begin{figure}[thb]
    \includegraphics[width=0.45\textwidth]{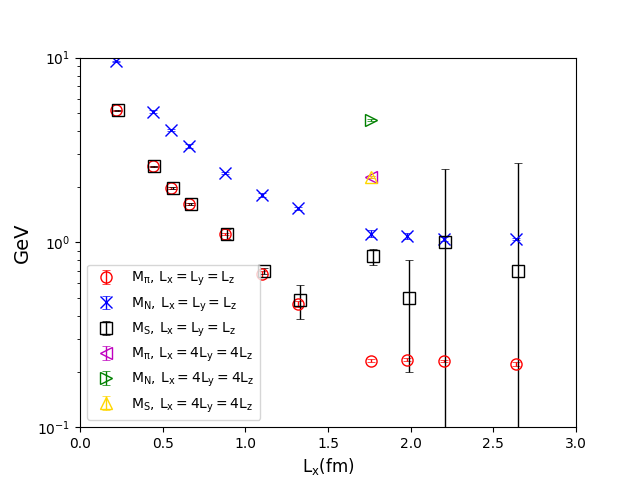}
    \caption{The non-singlet pion, $a_0$ and nucleon masses $m_\pi$, $m_{a0}$ and $m_N$ (upper panel), and mass difference $m_N-3/2m_{\pi}$ on the $L^3\times96$ lattice at $a\sim0.055$ \;fm}
    \label{fig:hadron_masses_L}
\end{figure}

The mass splitting between the $a_{0}$ and pion also serves as indicator of the chiral symmetry breaking. We determine $m_{a_{0}}$ using the same formula and fitting patterns as in the case of nucleon. As depicted in Fig.~\ref{fig:s_eff_masses_L} and Fig.~\ref{fig:hadron_masses_L}, while the mass degeneration the two mesons occur naturally of as $\mathcal{C}_{P}(t)=-\mathcal{C}_{S}(t)$ at $L < {\rm 0.7\;fm}$, the mass split are still small with $L\in(0.9,1.5)$ \;fm, even though obvious difference has been observed in the two point functions above $1.1$\;fm. $m_{a_{0}}$ increase to about 0.8 GeV at $L \sim 1.8$\;fm if we fit $\mathcal{C}_{S}(t)$ at small $t$, and then the uncertainty of $\mathcal{C}_{S}(t)$ increases rapidly with even larger $L$. This observation may indicate that the chiral phase transition using $L$ as the order parameter would be a crossover, similar to that using $T$ at finite temperature.

The results of hadron masses are summarized in Fig.~\ref{fig:hadron_masses_L}, showing $m_{\pi}$ and $m_{N}$ stabilization for $L>1.8$\;fm, suggesting minimal finite volume effects, whereas $m_{a_{0}}$ exhibits fluctuations due to complex $0^+$-state contamination. 

We also show an additional case on a $32\times8^2\times96$ lattice for comparison in Fig.~~\ref{fig:hadron_masses_L}. The hadron masses $m_{\pi, N, a_0}$ are quite close to those on the $8^3\times96$ lattice. It suggests that QCD properties would be dominated by the spatial direction with shorter extent in those cases.

\section{Summary}

We studied the hadron spectrum in the $\epsilon$-regime with near-physical quark masses with different $L$, using the clover valence fermion on the HISQ sea configurations at $a=$0.05 \;fm. The results show that the pion mass receives a huge $1/L$ correction at small $L$, leading to an increase in the corresponding $m_{\pi}L$ when $L\le 1.2$ \;fm, where the chiral symmetry breaking between the scalar and pseudoscalar correlation functions is effectively restored. At the same time, the mass gap $m_{N}-3m_{\pi}/2$ is always larger than 700 MeV in the entire $L\in(0.4,2.5)$ \;fm range we studied. Such an observation would suggest that the confinement of QCD can exist when the spontaneous chiral symmetry breaking is absent.

Since the valence clover fermion action we used breaks the chiral symmetry additively, it would be non-trivial to further investigate the origin of the restoration of the chiral symmetry and the remaining mass gap. The overlap fermion action would be more suitable for this task, and the study is ongoing. The mixed action effect on the above observations would also be a concern, and further study at multiple lattice spacings is also essential. In addition, the ensemble we used with the largest $L$ gives $m_{\pi}L\sim 3$, which is still smaller than the usual requirement $m_{\pi}L\sim 3.5$ which can suppress the finite volume effects efficiently. Thus, simulation with a larger $L$ would be helpful verify the infinite volume limit. 

\section*{Acknowledgement}
We thank the MILC collaborations for sharing their inputs of the gauge configurations, Carleton Detar, Keh-Fei Liu and Shoji Hashimoto for useful information and discussion. The calculations were performed using the MILC~\cite{MILC:2010pul,MILC:2012znn} and Chroma software suite~\cite{Edwards:2004sx} with QUDA~\cite{Clark:2009wm,Babich:2011np,Clark:2016rdz} through HIP programming model~\cite{Bi:2020wpt}. The numerical calculation were carried out on the ORISE Supercomputer, and HPC Cluster of ITP-CAS. This work is supported in part by NSFC grants No. 1229060, 1229062, 1229065 and 12047503, the science and education integration young faculty project of University of Chinese Academy of Sciences, the Strategic Priority Research Program of Chinese Academy of Sciences, Grant No.\ XDB34030303 and YSBR-101, and also a NSFC-DFG joint grant under Grant No.\ 12061131006 and SCHA 458/22.

\bibliography{ref}

\end{document}